\documentstyle[preprint,aps]{revtex}
\begin{document}
\draft
 
\title{FINAL STATE CHARGE EXCHANGE INTERACTIONS \\
IN THE $^{12}C(e,e'p)$ REACTION}
\author{S.~Jeschonnek$^{1}$, S.~Krewald$^{1}$ and A.~Szczurek$^{1,2}$}
\address{
$^{1}$ Institut f\"ur Kernphysik,
Forschungszentrum,
D-52425 J\"ulich, Germany\\
$^{2}$ Institute of Nuclear Physics,
ul. Radzikowskiego 152,
PL-31-342 Krak\'ow, Poland
}
\maketitle
 
\vspace{1cm}
\begin{center}
\end{center}
\begin{abstract}
The $^{12}C(e,e'p)$ reaction is analyzed in a model which
explicitly includes final state interactions due to the coupling
of the proton and neutron emission channels. We find that the effects
of the final state interactions due to charge exchange reactions
are important to get a good description of the symmetry properties of
the newly measured Mainz spectral functions. We discuss the possible role
the off-shell effects may play for the correct interpretation of
spectral functions at large positive missing momenta.
\end{abstract}
 
Recently, new data for the reaction $ ^{12}C(e, e'p) $
have been measured in parallel kinematics at the electron scattering
facility MAMI in Mainz with an energy of the incident electrons
of 855.1 MeV. The energy of the outgoing protons was varied from
$T_p$ = 82 MeV to $T_p$ = 118 MeV \cite{mainz}.
Previously, the same reaction has been measured at NIKHEF in Amsterdam
using electrons with energies ranging from 284 MeV to 481 MeV
and an energy of the outgoing protons of $T_p$ = 70 MeV \cite{nikhef}
and at Saclay with $T_p$ = 100 MeV protons \cite{saclay}.
 
The standard method to analyze the (e,e'p) reactions in quasielastic
kinematics is the distorted wave impulse approximation (DWIA) which
assumes that the interaction between the ejected nucleon and
the residual nucleus can be described by an optical potential
(see for instance \cite{boffi}).
It was found in Ref.\cite{nikhef} that the NIKHEF data for the
$^{12}C(e,e'p)$ reaction for negative missing
momenta are underestimated in standard DWIA approaches.
Moreover, it was found that the ratio of the transverse to
longitudinal response function exceeds the value expected
under the assumption of a quasifree reaction mechanism
\cite{steen,steen2}. The in-medium modification of the bound nucleons
has been proposed in a series of publications
\cite{nikhef},\cite{steen} - \cite{steen1}
as a possible resolution of these discrepancies.
On the other hand, final state interactions due to charge
exchange reactions $(e,e'n)(n,p)$ have been suggested as
an alternative explanation \cite{steen3,jsck}.
 
In this communication, we want to investigate whether
the new Mainz data corroborate the importance of final state
interactions and channel coupling
in the $(e,e'p)$ reaction. A detailed knowledge of
final state interactions in the kinematic regions probed by the
Mainz and the Amsterdam data is required in the analysis
of the forthcoming experimental data of quasi-elastic neutrino
scattering on carbon\cite{Garvey,musolf}.
Moreover, a good knowledge of the final state interactions
is mandatory to disentangle the effects of short range
correlations and final state interactions at higher
missing momenta.
 
We analyze the data within the framework of a continuum random
phase approximation which explicitly takes into account the
$^{12}C(e,e'p)$ and $^{12}C(e,e'n)$ decay channels.
Both direct and Pauli exchange diagrams are included.
The interaction is iterated up to infinite order.
The residual interaction has been derived from a one-boson
exchange potential \cite{nakayama}. Details of the model can
be found in Ref.\cite{jsck}.
 
The result of the calculations is compared with the new Mainz data
in Fig.1.
Since in the Mainz experiment different kinematical conditions
were employed for each data point, we performed the calculations
under precisely the same conditions.
The data can be grouped in three sets characterized by the
average scattering angles. To guide the eye, we have joined
the theoretical points of each set by a smooth curve. Near the
missing momentum $p_m$ = 100 MeV/c, one therefore can
distinguish two different theoretical curves.
 
Since the short range correlations in the ground state of $^{12}C$ are
not included in our model explicitly, the spectroscopic factor must be
used as an input parameter.  In our previous work \cite{jsck} we have
found that the spectroscopic factor obtained from the theoretical work
of Cohen and Kurath \cite{coku} allows a good description of the
NIKHEF data.  We find that a description of similar quality of the
Mainz data with this factor is not possible. An additional factor of
0.76 is needed to describe the Mainz data. We feel that the absolute
normalization of the data sets \cite{mainz} and \cite{nikhef,saclay}
does not agree and this issue must be clarified by further
experiments.  Now let us concentrate on the symmetry properties of the
data and on the effects of the final state interactions, disregarding
the problem of the absolute normalization.  A standard DWIA calculation
(dashed line) shows a relatively large asymmetry of the spectral
functions for positive and negative missing momenta. The 
final state interactions due to the coupling of the proton and
neutron emission channels reduce the asymmetry, however.
 
Both the NIKHEF and Mainz data sets show only a slight asymmetry.
The spectral function of Ref. \cite{mainz},  averaged over
the range of missing momenta from 77.5 MeV/c to 92.5 MeV/c,
is 34.0(GeV/c)$^{-3}$,
while the corresponding average of missing momenta ranging
from -92.5 to -107.5 MeV/c is 30.09 (GeV/c)$^{-3}$,
i.e. the maximum of the spectral function for positive missing
momenta is larger by a factor of 1.13 than the maximum for
negative missing momenta. The ratio of the maxima of the
spectral function for positive and for negative missing
momenta for the  data  of Ref. \cite{nikhef} is 1.14.
We get a comparable asymmetry of 1.10 in our approach,
whereas the mean field calculation leads to a ratio of 1.30.
 
In order to understand the dominant effects of FSI on the
asymmetry of the spectral function in Fig.2 we
present results when different final state interaction effects
are gradually taken into account. In this calculation for simplicity
the energy of the outgoing protons was fixed to $T_{p}$ = 75.3 MeV.
The  mean field calculation (dotted line) leads to a strong
asymmetry of the spectral function. This effect is predominantly
due to the real part of the optical potential. The imaginary part
causes only an overall damping of the spectral function with respect
to the plane wave impulse approximation.
The inclusion of the microscopic couplings between different
emission channels (solid line) leads to a partial restoration of the symmetry
of the spectral function with respect to $p_m$ = 0.
The relevance of the charge exchange reactions
$(e,e'p)(p,n)$ and $(e,e'n)(n,p)$ is demonstrated by switching off
the proton-neutron coupling (compare the dashed and solid curves).
The effect of the proton-neutron coupling is particularly large
near the maximum of the spectral function at the missing momentum
of $p_m$ = 100 MeV/c which corresponds to a momentum transfer
of $q \approx$ 280 MeV/c. For small momentum transfers the coupling
of the virtual photons to the neutron via the charge operator
vanishes. The strong final state interaction
$(e,e'p)(p,n)$ feeds the $(e,e'n)$ channel at the expense of
a depletion observed in the $(e,e'p)$ channel.
For large momentum transfers, the virtual photon couples mainly
via the magnetic current. The strength of the magnetic coupling
is approximately of the same size for protons and neutrons
and therefore the effect of the subsequent final state interaction
is less important. We wish to
emphasize in this context that, as discussed in Ref.\cite{jsck}, our
microscopic approach which explicitly includes realistic
nucleon-nucleon residual interactions substantially differs from
the simple mean field Lane potential applied in Ref.\cite{SBTM87}.
 
Let us come back now to the problem of the absolute normalization.
Here the difference of kinematical conditions can be of some
relevance. While at NIKHEF and Saclay the kinetic energy of the
knocked out proton was kept constant at $T_p$ = 70 MeV or $T_p$ = 100
MeV, respectively, and the momentum transfer $\vec q$ was varied, in
the Mainz experiment both the kinetic energy of the proton $T_p$ and
the momentum transfer $\vec q$ were varied.  It is known from the
nucleon-nucleus scattering phenomenology that the effective mean field
interaction is strongly energy dependent (see for example
\cite{BM67,Sch82}).  It has been shown in Ref.\cite{jsck} that this is
also a kinematical region sensitive to the final state interaction
effects caused by charge exchange reactions.  In distinction to
Ref.\cite{SBTM87}, in our model these FSI effects depend not only on
the energy of the outgoing nucleon but also on the transferred
momentum. Can these effects account for the disagreement of the
absolute normalization of the NIKHEF \cite{nikhef} and the Mainz
\cite{mainz} data? To answer this question in Fig.3 we show the
results of our full calculation for $T_{p}$ = 75.3 MeV (solid line),
$T_{p}$ = 90 MeV (dashed line) and $T_{p}$ = 108.5 MeV (dotted line),
covering the range of the Mainz kinematics. It can be seen that with
increasing kinetic energy, the spectral function is slightly shifted 
towards positive missing momenta.  This is a combined effect of the
energy dependence of the real and imaginary part of the effective
optical potential. As seen from Fig.3 the absolute value of the cross
section is only minimally affected.
 
Our model slightly overestimates both the NIKHEF and Saclay
experimental spectral functions at large positive missing momenta
\cite{jsck}.  It should be noted in this context that a similar strong
discrepancy between DWIA and experimental spectral functions has been
observed also in the NIKHEF experiment at rather low energy of the
outgoing protons \cite{nikhef} ($T_p$ = 40 MeV).  This may, however,
be sensitive to the details of the nucleon-nucleon interaction
used. In order to gauge the interaction used it would be very useful
to study experimentally not only $(e,e'p)$ but also $(e,e'n)$ spectral
functions. Here the coupling of the proton and neutron emission
channels may play even a more pronounced role \cite{jsck}.
 
The reader should note also that the discrepancy happens in the region
where the off-shell effects seem to be very strong.
In Fig.4 we present a ratio of the $e - p$
off-shell cross section calculated with the $cc1$ prescription
of de Forest \cite{deForest} to the standard on-shell Rosenbluth cross
section. This quantity seems to be a good measure of the off-shell
effects.  We show the ratio for the total (solid) cross sections as well
as for the longitudinal (dashed) and transverse (dotted) ones.
Because in the kinematical range appropriate for the Mainz experiment
the transverse part of the cross section is rather small, the ratio
of the total cross sections follows that for the longitudinal part.
As seen from the figure, the off-shell effects seem to be important
only at rather large positive missing momenta.
As there is no unique theoretical way to derive the $e-p$ off-shell
cross section (see detailed discussion in \cite{NPKO90}),
the interpretation of the data for positive missing
momenta (small transferred momentum) may be somewhat disturbed.
We wish to point out that in parallel kinematics the off-shell
effects in the current conserving $cc1$ prescription of
de Forest are independent of the electron beam energy as well as
of the energy of the outgoing particles. They depend dominantly
on the missing momentum. This strongly suggests that the off-shell
effects cannot be responsible for the difference between the Mainz
\cite{mainz} and NIKHEF \cite{nikhef} data.
 
To summarize, we have shown that 
neither the energy dependence of the final state interactions nor
the off-shell effects are able to resolve the question of
the spectroscopic factors, i.e. the effect of about 30$\%$ difference
in absolute normalization of the recent Mainz
\cite{mainz}, NIKHEF \cite{nikhef} and Saclay \cite{saclay}
spectral functions.
Our model leads to a satisfactory description
of the dependence of the spectral functions of Refs.
\cite{mainz} - \cite{saclay} on the missing momenta without
the introduction of a medium modification because the coupling
between proton and neutron emission channels is explicitly
taken into account. Experimental data on the $(e,e'n)$
reaction would provide a sensitive test of our model.

\vspace{1cm}
 
{\bf Acknowledgments:}
We are indebted to J.Speth for pointing out the problem and discussion.
We are also grateful to E.Offermann for supplying
us with the precise kinematics of the Mainz experiment and
J.Friedrich and G. van der Steenhoven for stimulating
correspondence.

 
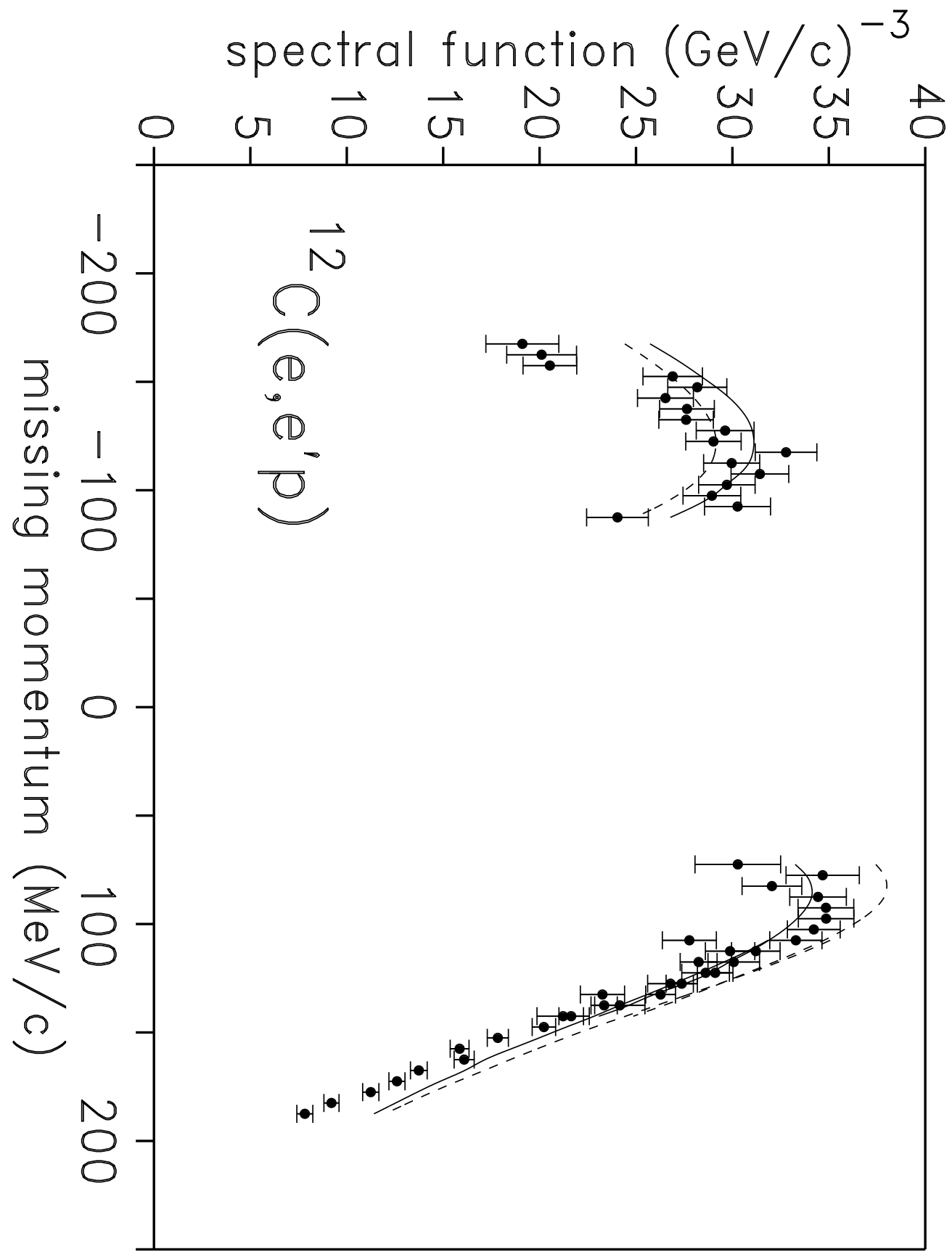
\begin{figure}
\caption{The spectral function in units of (GeV/c)$^{-3}$
is shown as a function of the missing momentum $ p_m $, given in
MeV/c. The result of a DWIA analysis is shown by the dashed line.
The effects of final state interactions due to charge exchange
reactions are given by the solid line.
Since each data point of Ref.\protect \cite{mainz} was taken for
slightly different kinematical conditions, the theoretical
calculations have been performed  under exactly the same conditions.
The theoretical points have been joined by three smooth curves to
guide the eye. }
\end{figure}
 
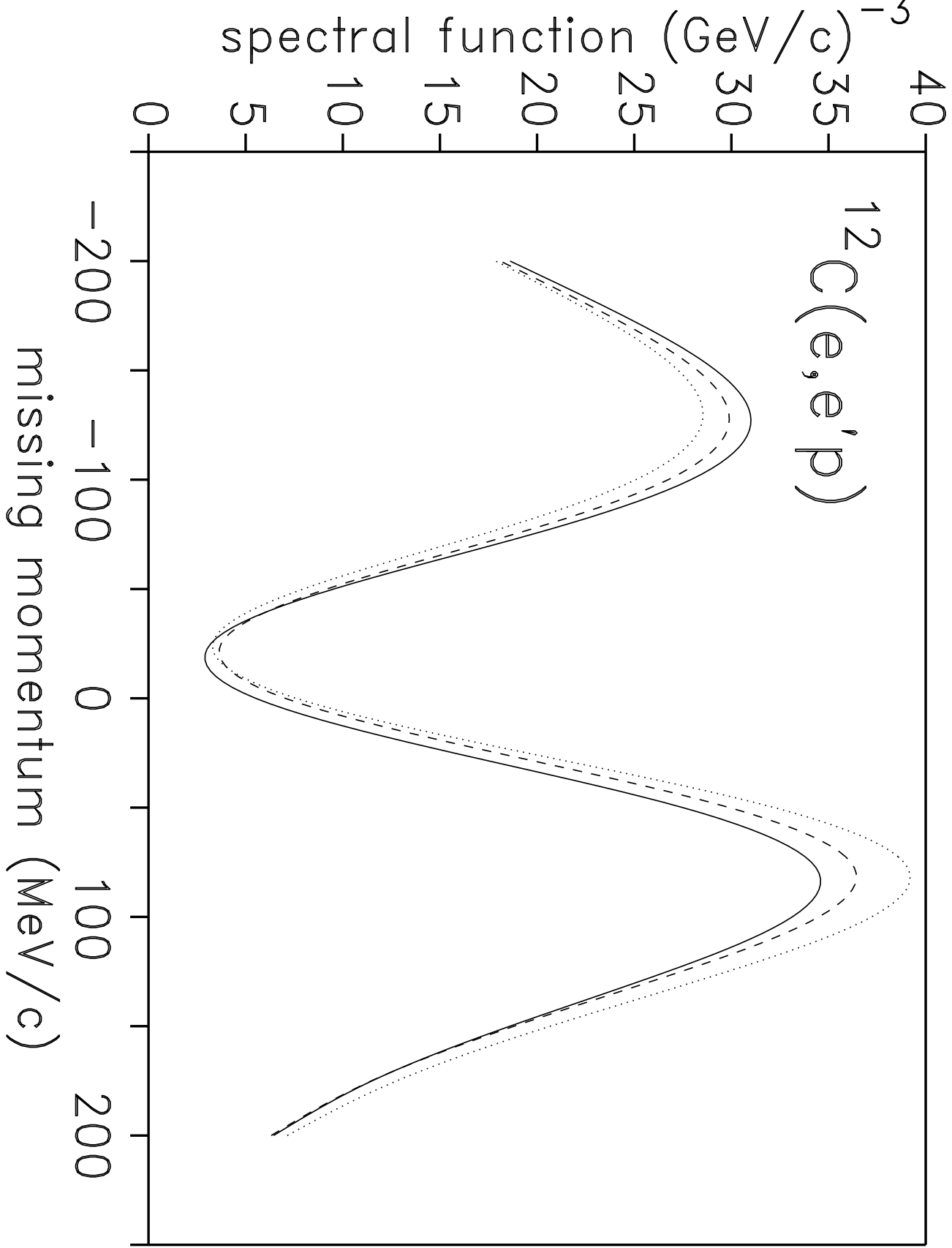
\begin{figure}
\caption{The influence of FSI effects on the asymmetry of the spectral
function is shown. We compare results obtained in the mean-field
approximation (dotted line), including microscopic coupling between
the proton emission channels only (dashed line) and the full
calculation including also the coupling of the proton and neutron
channels (solid line). }
\end{figure}
 
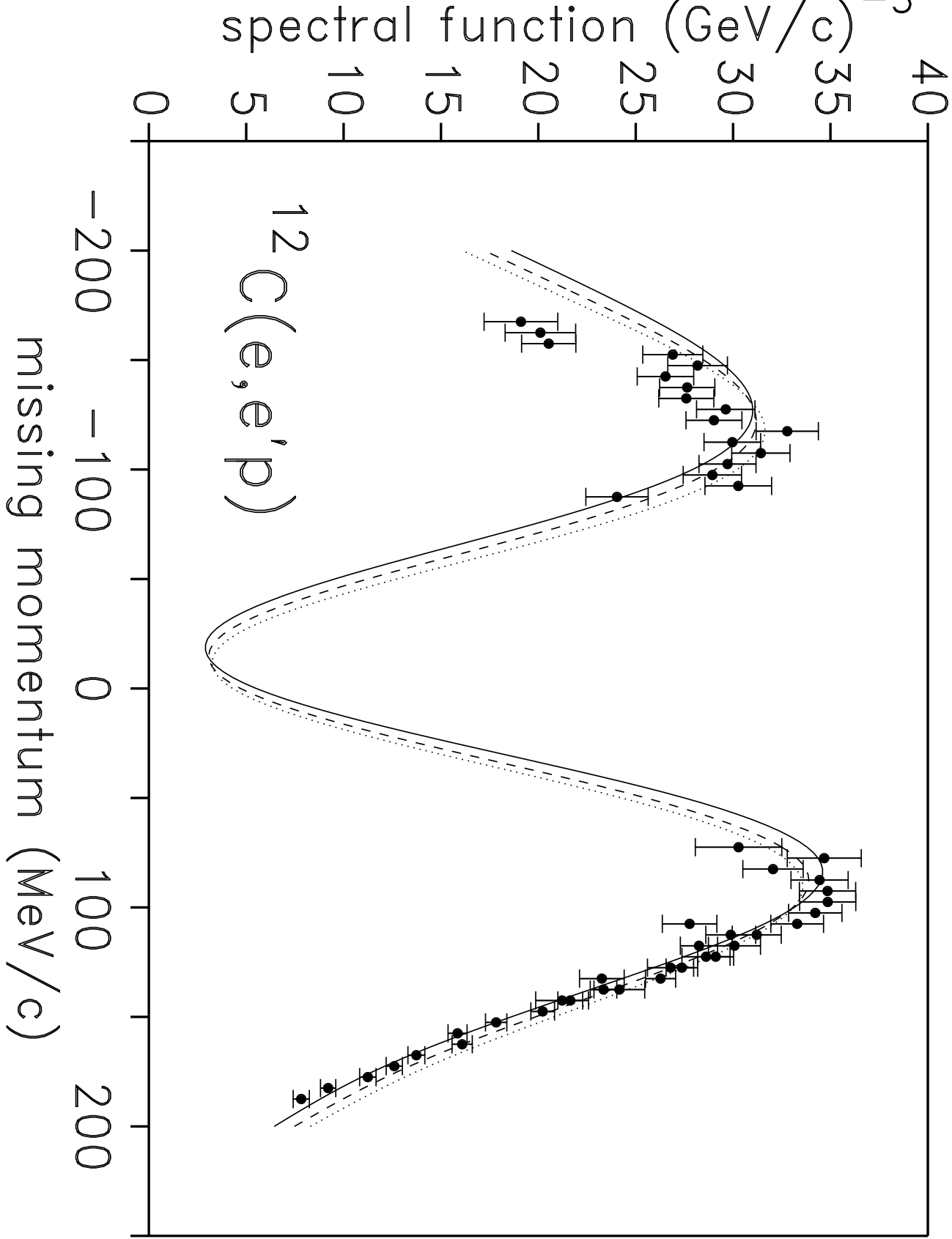
\begin{figure}
\caption{The spectral function in units of (GeV/c)$^{-3}$
is shown as a function of the missing momentum $ p_m $, given in
MeV/c for three different kinematical conditions: the energies of the
emitted protons are kept fixed at $T_p$ = 75.3 MeV (dotted line),
$T_p$ = 90.0 MeV (dashed line), and
$T_p$ =108.5 MeV (solid  line), respectively. }
\end{figure}
 
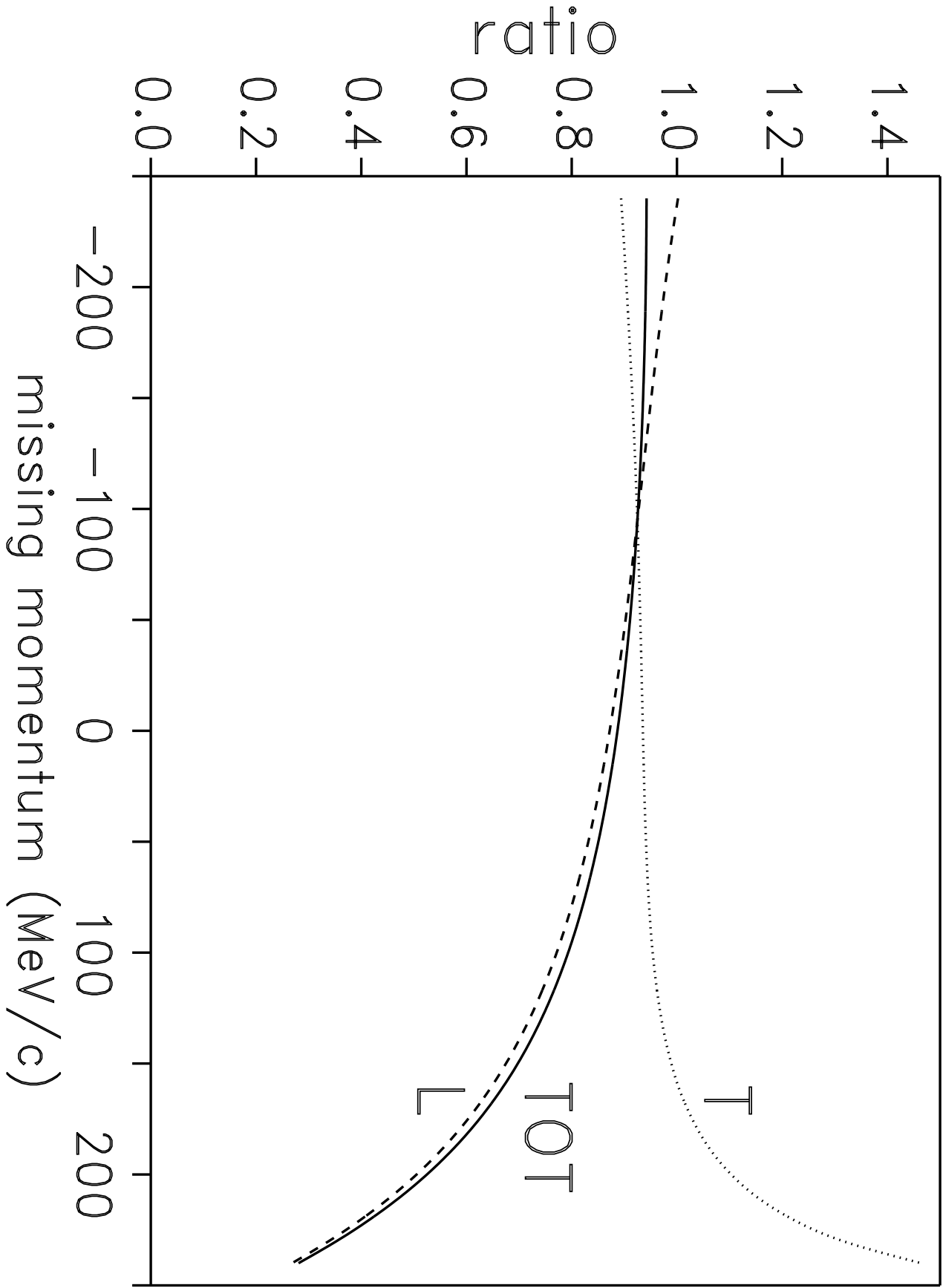
\begin{figure}
\caption{
The ratio of the off-shell to the on-shell $e-p$
cross section in the (anti)parallel kinematics as a function
of the missing momentum. }
\end{figure}


\begin{thebibliography}{99}
 
\bibitem{mainz}
K.I.Blomqvist et al., {\sl Z.Phys.A} {\bf 351}, 353 (1995).
 
\bibitem{nikhef}
G. van der Steenhoven, H.P.Blok, E.Jans, M.deJong, L.Lapikas,
E.N.M.Quint, and P.K.A. deWitt Huberts,
{\sl Nucl.Phys.}{\bf A480}, 547 (1988).
 
\bibitem{saclay}
M.Bernheim et al., {\sl Nucl.Phys.}{\bf A375}, 381 (1982).
 
\bibitem{boffi}
S.Boffi, C.Giusti, and F.D.Pacati, {\sl Phys.Rep.}{\bf 226}, 1 (1993).

\bibitem{steen}
G. van der Steenhoven, E.Jans, L.Lapikas,
E.N.M.Quint, and P.K.A. deWitt Huberts,
{\sl Phys.Rev.Lett.}{\bf 57}, 182 (1986).
 
\bibitem{steen2}
G. van der Steenhoven et al.,
{\sl Phys.Rev.Lett.}{\bf 58}, 1727 (1987).


\bibitem{nikhef1}
G. van der Steenhoven et al.,
{\sl Nucl.Phys.}{\bf A484}, 445 (1988).
 
 
\bibitem{steen1}
D.G.Ireland and G.van der Steenhoven
{\sl Phys.Rev.C.}{\bf 49}, 2182 (1994).

\bibitem{steen3}
G.van der Steenhoven et al., {\sl Nucl.Phys.}{\bf A527}, 17c (1991).

\bibitem{jsck}
S.Jeschonnek, A.Szczurek, G.Co', and S.Krewald,
{\sl Nucl.Phys.}{\bf A570}, 599 (1994).
 

\bibitem{Garvey} G.Garvey, E.Kolbe, K.Langanke, and S.Krewald,
 {\sl Phys.Rev.}{\bf C48}, 1919 (1993).

\bibitem{musolf}
M.J.Musolf, T.W.Donnelly, J.Dubach, S.J.Pollock, S.Kowalski,
and E.J.Beise, {\sl Phys.Rep.} {\bf 239}, 1 (1994).


\bibitem{nakayama} K.Nakayama, S.Krewald, J.Speth and W.G.Love,
{\sl Nucl.Phys.}{\bf A431}, 419 (1984).
 

\bibitem{coku} S.Cohen and D.Kurath,
 {\sl Nucl.Phys.}{\bf 73}, 1 (1965)
;{\bf A101}, 1 (1967).

\bibitem{SBTM87} G.van der Steenhoven, H.P.Blok, M.Thies and P.Mulders,
{\sl Phys.Lett.}{\bf B191}, 227 (1987).
 
\bibitem{BM67} A.Bohr, B.R.Mottelson,
Nuclear Structure, W.A.Benjamin, Inc., New York, 1967.
 
\bibitem{Sch82} P.Schwandt et al.,
{\sl Phys.Rev.}{\bf C26}, 55 (1982).


\bibitem{deForest} T.de Forest, {\sl Nucl.Phys.}{\bf A392}, 232 (1983).


\bibitem{NPKO90} H.W.L.Naus, S.J.Pollock, J.H.Koch and U.Oelfke,
{\sl Nucl.Phys.}{\bf A509}, 717 (1990).
   
 


 

\end{thebibliography}
\end{document}